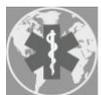

*Article*

# A Bayesian Downscaler Model to Estimate Daily PM2.5 levels in the Continental US

**Yikai Wang** [1], **Xuefei Hu** [2], **Howard Chang** [1], **Lance Waller** [1], **Jessica Belle** [1], **Yang Liu** [2,*]

[1] Department of Biostatistics and Bioinformatics, Rollins School of Public Health, Emory University, Atlanta, GA 30322, USA
[2] Department of Environmental Health, Rollins School of Public Health, Emory University, Atlanta, GA 30322, USA
* Correspondence: yang.liu@emory.edu

**Abstract:** There has been growing interest in extending the coverage of ground PM2.5 monitoring networks based on satellite remote sensing data. With broad spatial and temporal coverage, satellite based monitoring network has a strong potential to complement the ground monitor system in terms of the spatial-temporal availability of the air quality data. However, most existing calibration models focused on a relatively small spatial domain and cannot be generalized to national-wise study. In this paper, we proposed a statistically reliable and interpretable national modeling framework based on Bayesian downscaling methods with the application to the calibration of the daily ground PM2.5 concentrations across the Continental U.S. using satellite-retrieved aerosol optical depth (AOD) and other ancillary predictors in 2011. Our approach flexibly models the PM2.5 versus AOD and the potential related geographical factors varying across the climate regions and yields spatial and temporal specific parameters to enhance the model interpretability. Moreover, our model accurately predicted the national PM2.5 with a R2 at 70% and generates reliable annual and seasonal PM2.5 concentration maps with its SD. Overall, this modeling framework can be applied to the national scale PM2.5 exposure assessments and also quantify the prediction errors.

## 1. Introduction

Particulate air pollution has become a major environmental and public health concern worldwide in recent years. Particularly, particulate matter with aerodynamic diameter <= 2.5 um (PM2.5) is shown to have a strong association with various adverse health outcomes such as the increased mortality and morbidity, aggravated respiratory and cardiovascular symptoms (1). Ambient PM2.5 is either directly emitted from various anthropogenic and biogenic sources or generated in the atmosphere from complex photochemical reactions (2). Consequently, PM2.5 concentrations vary in space and time at sub-kilometer to continental scales (3). Thus, it is important to accurately assess the population exposure of PM2.5. However, in interest of reducing cost, PM2.5 monitors are usually sparsely distributed and tend to be concentrated among the urban areas and most PM2.5 monitors operated by US EPA and the IMPROVE network only operate on a one-in-three-days or one-in-six-days schedule, leaving significant temporal gaps. Due to these spatial and temporal limitations, it is difficult for current PM2.5 networks to provide sufficient data to fully assess PM2.5 for health effect studies and it could lead to biased results for some key scientific questions.

One emerging solution to these problems is spatial models driven by remotely sensed particle properties from the satellite platform as well as gridded meteorological and land use information. The most robust and widely used satellite parameter is the aerosol optical depth (AOD), which measures the overall particle light extinction caused by airborne particles in the atmospheric column.

Many previous studies have shown that AOD has a strong positive association with PM2.5. In addition to AOD, previous studies have shown that the meteorological and land use information are important factors to predict ground level concentration of PM2.5 and the relationship between AOD and PM2.5 (4-6). All these properties between AOD and PM2.5 make it possible to develop the statistical methods to calibrate the PM2.5 using AOD and other geographical factors. Over the past decade, various MODIS-driven PM2.5 exposure models have been developed, from relatively simple linear regressions (7) to complex multi-level spatial models (8) and Bayesian hierarchical models (9). Bayesian hierarchical models have more flexibility in modeling the complex temporal and spatial pattern of PM2.5 and compared with other spatial models based on mixed effect terms, one major advantage of Bayesian model is its underlying nature to quantify the prediction uncertainty through MCMC algorithm which is crucial for scientific research. Therefore, in this session, we extend the Bayesian model proposed by Chang et al. (2014) into a national Bayesian model to study the PM2.5 under a national domain.

So far, most satellite-driven PM2.5 exposure models have been developed at the urban to regional scales in order to support health effect studies in the specific regions (5, 9-14). High-performance national scale PM2.5 exposure models are still limited partially because of the high-computational demand in order to make national PM2.5 prediction surfaces. A couple of national-scale studies involve machine learning methods (15, 16). Di et al. (2016) developed a neural network approach, incorporated with convolutional layers to account for spatiotemporal autocorrelation, to predict PM2.5 concentrations in the continental U.S. from 2000 to 2012. Hu et al. (2017) developed a random forest model with ~40 predictors to predict PM2.5 exposure in the conterminous U.S. in 2011. These emerging methods can provide relatively high predication accuracy but offer little insight into how different predictors behave across such large domains. For example, random forests only provide an importance value for each predictor to indicate which predictor is more important in training process. Both neural networks and random forests do not provide quantification of uncertainties in prediction and parameter estimation. These methods also cannot provide straightforward estimates of the model prediction errors. On the other hand, statistical models provide a balance between the model predication accuracy and the ability for interpretation and serves as the most reliable and commonly used approaches in calibrating the PM2.5. For example (12) proposed a mixed effect model with random temporal intercept and slope on AOD to evaluate the time-varying effects. This type of model assumes that the temporal random effect based model requires the independence assumption between different days which is generally not practical and they have limited power to make prediction out of the temporal domain. They also failed to adjust the spatial variability existed in the large spatial domain and can provide biased results. Similar for the hierarchical model provided by (5, 10, 11), it is tricky to quantify the uncertainties in prediction or parameter estimation based on such models which limits its power to real application. Thus, all these models are not directly applicable to the national-wise domain.

Chang et al. (9) reported a Bayesian downscaling model which adopted the Gaussian spatial process to incorporate the spatial correlation into the model which increases the power to borrow information across the neighborhoods, through which the challenge of spatial misalignment between the point-referenced monitoring measurements and the gridded areal AOD data can be solved, and it models the conditional correlation between adjacent observed days which allows us to estimate the random effects on the day without PM2.5 measurements. In addition, this model adopts a full Bayesian approach where the model uncertainty can be obtained easily. However, this model is only applicable to a small spatial domain for three reasons. First, it assumes that the temporal correlation structure is constant across the whole spatial domain but based on our study this is not realistic in such a large spatial domain. Similarly it assumes that the spatial correlation structure is constant across the whole year which is not realistic. Second, the original model is not flexible enough to capture the huge spatial variability existed in the national domain. For example, it assumes a constant effect of land use effect across different states which failed to consider the localized difference which is one of the goals of a national-wise study. Finally, directly generalizing the original model computationally expensive because the spatial correlation matrix is of high dimensions and very sparse.

In this paper, we enhanced and expanded the original Bayesian downscaler to the entire continental U.S. We developed a regional and temporal specific Bayesian downscaling approach to gain more flexibility. Our model incorporates AOD data, meteorological fields and land-use variables to estimate daily ground-level PM2.5 concentrations over the conterminous United States for the year 2011. The estimated regional and temporal specific parameters are scientifically meaningful and the prediction accuracy is evaluated through the general and spatial cross validation frameworks. At last, our model predicts the daily averaged PM2.5 concentrations across the entire continental U.S. and also the prediction uncertainty maps.

**2. Data and Methods**

*2.1 Data Collection*

The 24-hour averaged PM2.5 measurements for 2011 were downloaded from the U.S. Environmental Protection Agency (EPA)'s Air Quality System Technology Transfer Network (http://www.epa.gov/ttn/airs/airsaqs/). Collection 6 level 2 Aqua MODIS retrievals at a spatial resolution of 10 km were regridded to the 12 x 12 km2 Community Muti-Scale Air Quality (CMAQ) modeling system, and we calculated AOD averages using AOD retrievals from the combined deep-blue and dark-target parameter. Meteorological fields were obtained from the North American Regional Reanalysis (NARR) (http://www.emc.ncep.noaa.gov/mmb/rreanl/) with a spatial resolution of ~32 km and a temporal resolution of three hours and the North American Land Data Assimilation System Phase 2 (NLDAS-2) (http://www.emc.ncep.noaa.gov/mmb/rreanl/) at a spatial resolution of ~13 km and a temporal resolution of one hour. Elevation data at a spatial resolution of ~30 m were downloaded from the National Elevation Dataset (http://ned.usgs.gov). Road data were extracted from ESRI StreetMap USA. Percentage forest cover at a spatial resolution of ~30 m were extracted from the 2011 Landsat-derived land cover map downloaded from the National Land Cover Database (NLCD) (http://www.mrlc.gov). Primary PM2.5 emissions were obtained from the 2011 EPA National Emissions Inventory (NEI) facility emissions report (https://www.epa.gov/air-emissions-inventories/2011-national-emissions-inventory-nei-data).

*2.2 Climate Regions and Temporal Domains*

To improve computational efficiency, we divided the CONUS into nine NOAA-defined climate regions (Karl and Koss(1984)), which include Northeast, Southeast, South, Ohio Valley (Central),Upper Midwest(East North Central),Northern Rockies and Plains (West North Central), Southwest, Northwest and West, Figure 1. After examining aerosol light extinction measurements in various regions of the world, (3) reported that the typical mesoscale variability of lower-tropospheric particles ranges between 40 – 400 km Therefore, by dividing our national domain into nine multi-state regions we are still able to sufficiently capture the spatial and temporal correlations of ground-level PM2.5. We added a 100 km buffer to each climate region, and averaged overlapping predictions from neighboring regions to generate a smooth national PM2.5 concentration surface. In addition, the spatial pattern varies significantly across the year. To flexibly model this temporally evolving pattern while maintaining the simplicity, we divided the year of 2011 into three 4-month temporal periods and develop a Bayesian Downscaling model in each period. Since the typical PM2.5 residence time in the boundary layer ranges from a couple of days to two weeks, this treatment had minimum impact on model performance.

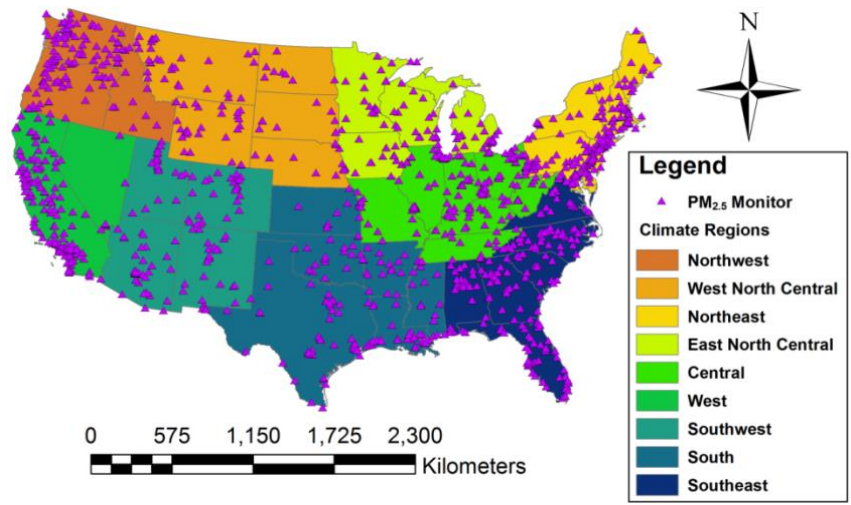

**Figure 1.** The Nine Climate Region and the Spatial Location of the Monitors.

*2.3 National Bayesian Downscaling Model*

For each regional and temporal sub-domain, we adopted the basic framework of the Bayesian Downscaling model proposed by Chang et al. (2013). In this model, let PM(s,t) denote the PM2.5 concentration at location s and day t, where s can be viewed as the unique spatial coordinates. Similarly, let AOD(s,t) denote the AOD measurement at the grid cell containing the monitor s and day t. For one specific climate region reg, a function of s, and temporal domain, the first level model between AOD and PM2.5 is given as:

$$PM(s,t) = \alpha_0(s,t) + \alpha_1(s,t)\, AOD(s,t) + \gamma_{reg,tem}(s,t)\, Z(s,t) + \varepsilon(s,t), \quad (1)$$

where $\alpha_0(s,t)$ and $\alpha_1(s,t)$ are the day-specific and location-specific random intercept and slope and the residual error $\varepsilon(s,t)$ is assumed to be independently normal with mean zero and regional- and temporal-specific variance $\sigma_{reg,tem}^2$. $Z(s,t)$ represents for the covariates having a constant association with PM2.5 where $\gamma_{reg,tem}$ represents for the regional- and temporal-specific fixed effect between $Z(s,t)$ and PM(s,t). Here Z(s,t) includes fire, forest coverage, emission, RH, temperature, wind speeds, major roadway length, hpbl and the interaction between AOD and temperature.

The spatial-temporal random effects $\alpha_0(s,t)$ and $\alpha_1(s,t)$ are specified using additive setting. For clarity, we present the model setting for one specific region and temporal domain: $\alpha_i(s,t) = \beta_i(s) + \beta_i(t)$, i = 0,1, where $\beta_i(s)$ and $\beta_i(t)$ are independent spatial and temporal effects. The spatial effects are modeled using a latent structure of two independent spatial Gaussian process $W_1(s)$ and $W_2(s)$, where $\beta_0(s) = c_1 W_1(s)$ and $\beta_1(s) = c_2 W_1(s) + c_3 W_2(s)$ and the covariance function of $W_i(s)$ for each region is given by exponential function multiplied by a tapering function. The regional-specific temporal effects $\beta_0(t)$ and $\beta_1(t)$ are modeled as two independent daily time series using a first-order random walk which can be defined through the conditional distribution of a particular day given all other days.

*2.4 Model Fitting and Prediction*

Model fitting was carried out using Markov Chain Monte Carlo (MCMC) techniques. Details of the MCMC algorithms and the prior settings can be found in online supplementary materials. Prediction performance was evaluated using two different cross-validation methods: fully random cross-validation (R-CV) and spatial cross-validation (S-CV). In R-CV, we randomly split the data into 10 folds and fit the model using 9 folds and evaluate the fitted model using the remaining fold, which can be used to evaluate the overall prediction ability of our approach. The S-CV is similar to R-CV except that rather than randomly splitting the data, we split the data based on its spatial location. The S-CV results were used to evaluate the ability in spatial extrapolation. In addition, through the MCMC approach, we quantified the prediction uncertainty, i.e. interval estimates. We also calculated the prediction statistics by comparing the predicted PM2.5 measurements with the observations, which include root mean squared error (RMSE), 90% posterior interval (PI) length and its empirical coverage probability and linear coefficient of determination R2 value. All analyses were carried out in R version 3.2.3.

## 3. Results

### 3.1 Data Description and Summary

The histograms of all variables are illustrated in Figure 1 which shows that all the variables are approximately unimodal and log-normal distributed. Log-transformation will be conducted to Fire, Emission and Road for the following analysis, and z-transformation is also conducted to all variables except for PM2.5 and AOD to remove the co-linearity between covariates and make the scale to be comparable. The annual mean PM2.5 concentration for all monitors is 9.88 μg/m3 with a SD at 6.17 μg/m3. The overall mean of AOD is 0.14 with a SD at 0.15. Specifically, the region-specific descriptive statistics for PM2.5 and AOD are summarized in Table 1. The number of records, monitors, days and the percentage of missing for each region are summarized in Table 2. Among 9 regions, OhioValley has the highest mean PM2.5 concentration at 11.29 μg/m3 and it has most records (18642) and monitors (361).

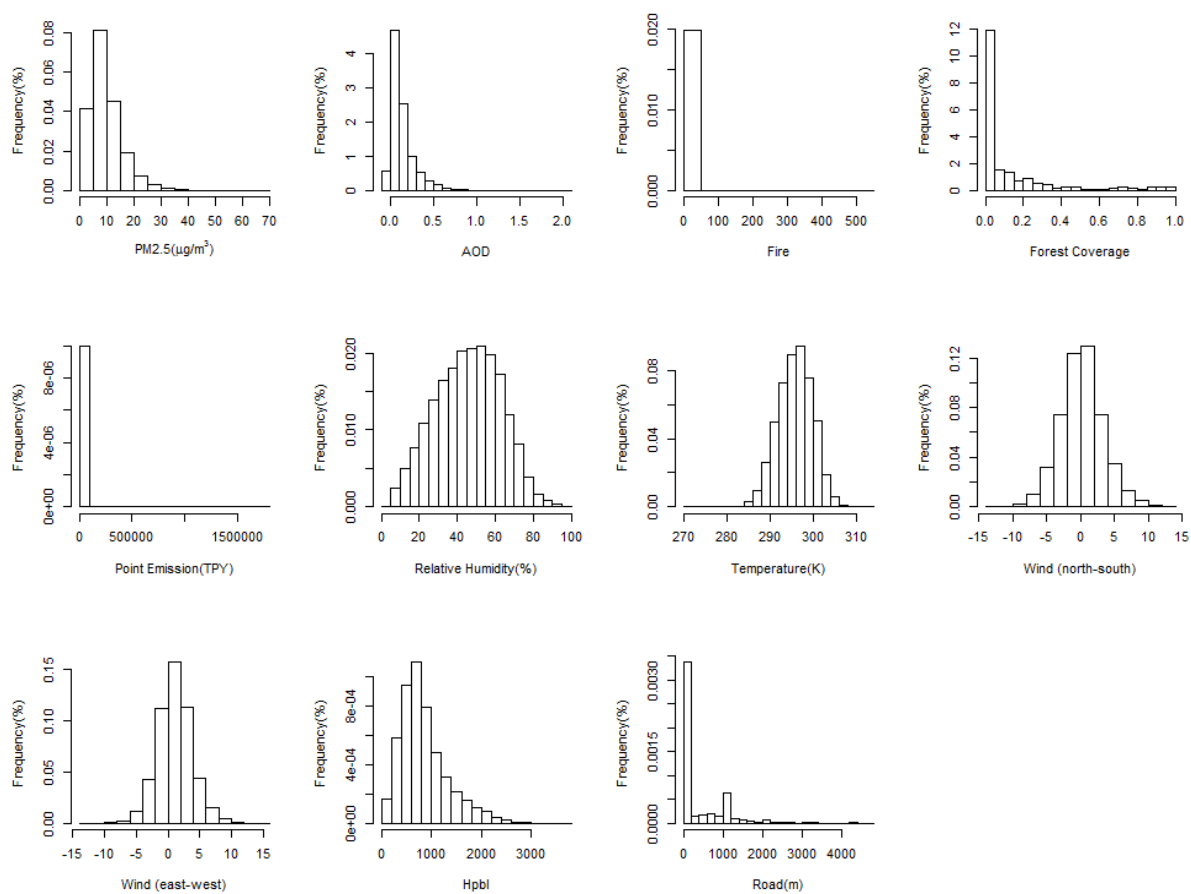

**Figure 2.** Histograms of the dependent and independent variables.

**Table 1.** Descriptive Statistics for PM2.5 and AOD.

| Regions   | PM2.5 (SD)   | AOD (SD)    |
|-----------|--------------|-------------|
| West      | 10.72 (7.17) | 0.10 (0.12) |
| Northwest | 6.23 (4.05)  | 0.12 (0.11) |
| Southwest | 7.40 (4.75)  | 0.10 (0.11) |
| NorthernRo | 7.40 (4.11) | 0.12 (0.13) |
| UpperMidwe | 10.33 (5.87) | 0.18 (0.17) |
| South | 10.17 (5.09) | 0.13 (0.15) |
| Southeast | 10.83 (5.34) | 0.15 (0.17) |
| OhioValley | 11.29 (5.79) | 0.17 (0.17) |
| Northeast | 10.68 (6.10) | 0.19 (0.19) |

**Table 2. Regional-specific Counts and Missing Pattern.**

| Regions | # of Records | # of Days | # of Monitors | % of Missing |
|---|---|---|---|---|
| West | 17096 | 356 | 159 | 0.30 |
| Northwest | 9486 | 295 | 170 | 0.19 |
| Southwest | 9567 | 363 | 138 | 0.19 |
| NorthernRo | 7463 | 328 | 150 | 0.15 |
| UpperMidwe | 6208 | 304 | 145 | 0.14 |
| South | 15899 | 364 | 189 | 0.23 |
| Southeast | 17525 | 361 | 257 | 0.19 |
| OhilValley | 18642 | 354 | 361 | 0.15 |
| Northeast | 8913 | 302 | 238 | 0.12 |

*3.2 Regional and Temporal Varying Geographical Associations*

In this section, we will explore the different significant patterns across nine climate regions and three temporal domains revealed by the significant parameters in National Bayesian downscaling model. First of all, the national Bayesian downscaling model fits well across different regions and temporal domains in terms of the R2 and model slope shown in Table 3 where the northeastern part of America tends to have highest performance of model fitting. The climate regions with the highest set of R2 are the Upper-Midwest (0.85) and Northeast (0.84) climate regions. Furthermore, the slopes in these two climate regions are consistently higher than 0.95 across all time domains indicating that there is no systematic bias in the model fitting. On the other hand, the model tends to have lower R2 in the southern and northwestern America areas, where the annual R2 for South climate region is 0.64 and the annual R2 for Northwest climate region is 0.63.

Table 3 presents all the significant geographical and meteorological factors (p value < 0.05), which exhibits substantial inter-region difference among the regional models. First of all, AOD is the most important covariate in this model which is significant for most regions and temporal domains. Specifically, we notice that the effect of AOD on PM2.5 is weaker in the second temporal domain (May - August) than other months and this pattern is commonly observed in all climate regions after we condition the temperature to be the average level in the specific spatial and temporal domain. Furthermore, Fire, RH, TMP and the interaction between AOD and TMP are significant across all regions. Other covariates including forest coverage, emission, wind speed, Hpbl and road length vary across regions and temporal domains. Specifically, the forest coverage is significant factor in explaining the pattern of PM2.5 in the western areas like West, Northwest and Southwest climate regions across the whole year, but is not significant in the West North Central climate region at all. Furthermore, emission is not significant in most regions but in the West climate region emission significantly explains the variability of PM2.5. On the other side, HPBL has a significantly negative association with PM2.5 in the West region but is not significant in the northern areas like Northwest and Northeast climate region.

Table 2. Regional-specific Counts and Missing Pattern.

| Region | Temporal | AOD | Fire | Forest | Emission | RH | TMP | Vgrd | Ugrd | Hpbl | Road | AOD*TMP | R2 | Slope |
|---|---|---|---|---|---|---|---|---|---|---|---|---|---|---|
| West | 1 | 21.2(5.9) | | -0.7(0.3) | 0.6(0.2) | 2.5(0.2) | 2.4(0.4) | 1.2(0.2) | | -0.2(0.1) | | 9.8(3.4) | 0.65 | 0.88 |
| | 2 | 4.1(1) | | -0.8(0.3) | 0.4(0.1) | 0.7(0.1) | 2.7(0.2) | | | -0.1(0.1) | | | 0.77 | 0.94 |
| | 3 | 31.2(5.2) | 0.2(0.1) | -1.9(0.3) | 0.8(0.3) | 0.6(0.1) | | 1.4(0.1) | -0.4(0.1) | -0.7(0.1) | | -8.5(2.5) | 0.72 | 0.91 |
| Northwest | 1 | | | -1.5(0.5) | -0.7(0.3) | -1.9(0.5) | | | | | | | 0.57 | 0.84 |
| | 2 | 5.4(1.1) | 0.1(0) | -0.2(0.1) | | 0.4(0.1) | 1.5(0.2) | | | | | 4.4(1) | 0.62 | 0.92 |
| | 3 | 25.4(3.8) | 0.4(0.1) | -0.4(0.2) | | | 1.2(0.4) | -0.4(0.1) | | | | 14.7(2.1) | 0.69 | 0.9 |
| Southwest | 1 | 10.6(5) | 0.3(0.1) | -0.7(0.2) | | 0.5(0.2) | 2.4(0.3) | 0.5(0.1) | | | | -11(2.8) | 0.69 | 0.89 |
| | 2 | 5.5(1.8) | | -0.3(0.1) | | 0.4(0.2) | 3.5(0.3) | 0.3(0.1) | 0.6(0.1) | | | | 0.6 | 0.88 |
| | 3 | 18.8(4.5) | | -0.5(0.2) | | | 0.7(0.2) | | -0.2(0.1) | -0.3(0.1) | | | 0.68 | 0.9 |
| NorthernRo | 1 | | 0.4(0.1) | | | 1.2(0.3) | | 0.7(0.2) | | | | | 0.82 | 0.95 |
| | 2 | 4.4(1.5) | 0.3(0.1) | | | 0.3(0.1) | 2.4(0.2) | 0.4(0.1) | | | | 3.1(1) | 0.67 | 0.92 |
| | 3 | 11.1(2.1) | 0.3(0.1) | | | | 2.1(0.2) | | -0.6(0.1) | -0.4(0.1) | | | 0.73 | 0.92 |
| UpperMidwe | 1 | | | | | 0.5(0.3) | | 1.3(0.2) | | -0.4(0.2) | | | 0.79 | 0.95 |
| | 2 | 4.4(1.7) | 0.3(0.1) | -0.6(0.2) | | 0.9(0.1) | 2.7(0.3) | 1(0.1) | -0.3(0.1) | | | 3.7(1) | 0.82 | 0.95 |
| | 3 | 9.5(3) | 0.4(0.1) | -0.6(0.2) | 0.3(0.1) | | 2.5(0.2) | 0.4(0.1) | -0.3(0.1) | -0.2(0.1) | | | 0.85 | 0.96 |
| South | 1 | 13.1(2.2) | 0.5(0) | -0.3(0.1) | | | 1.4(0.2) | 0.3(0.1) | -0.2(0.1) | | | | 0.59 | 0.91 |
| | 2 | | 0.2(0.1) | | | 0.5(0.1) | 4.2(0.3) | -0.2(0.1) | 0.2(0.1) | | 0.3(0.1) | 4.5(1.1) | 0.67 | 0.94 |
| | 3 | 14.7(1.9) | 0.3(0) | -0.7(0.1) | | -0.2(0.1) | 1(0.2) | 0.3(0.1) | -0.4(0.1) | -0.4(0.1) | 0.3(0.1) | | 0.65 | 0.93 |
| Southeast | 1 | 15.1(1.9) | 0.3(0) | -0.3(0.1) | | -0.3(0.1) | 0.8(0.2) | 0.5(0.1) | | | | 4.6(0.9) | 0.68 | 0.94 |
| | 2 | 4.6(1.6) | 0.1(0) | | | 1.7(0.2) | 7(0.4) | -0.6(0.1) | 0.2(0.1) | | | 6.1(1.1) | 0.74 | 0.95 |
| | 3 | 11.1(1.4) | 0.3(0) | -0.6(0.1) | | -0.7(0.1) | 0.8(0.2) | 0.7(0.1) | | -0.3(0.1) | | 6.9(1.1) | 0.69 | 0.94 |
| OhilValley | 1 | 21.2(2.9) | 0.7(0) | -0.5(0.1) |  | 0.4(0.1) | 0.7(0.2) | 0.7(0.1) | -0.3(0.1) |  |  | 5.5(1) | 0.68 | 0.94 |
|  | 2 | 5.7(1.3) |  |  |  | 2.2(0.1) | 5.5(0.3) | 0.3(0.1) |  |  | 0.2(0.1) | 2.9(0.7) | 0.74 | 0.95 |
|  | 3 | 14.4(5.2) | 0.3(0.1) | -0.8(0.1) |  |  | 1.7(0.2) | 0.5(0) | -0.3(0) | -0.3(0.1) |  | 3.3(1.3) | 0.77 | 0.95 |
| Northeast | 1 | 10.6(2.8) | 0.4(0.1) |  |  |  |  | 0.9(0.1) | -0.4(0.1) |  |  |  | 0.8 | 0.95 |
|  | 2 |  | -0.2(0.1) |  |  | 1.2(0.2) | 6.4(0.4) | -0.4(0.1) |  |  |  | 8.5(1.1) | 0.84 | 0.96 |
|  | 3 | 31(2.5) | 1.5(0.2) | -0.8(0.2) |  | 1.8(0.2) | 1.4(0.4) |  |  |  |  | 27.5(2) | 0.8 | 0.95 |

*All predictors are significant at $\alpha$ = 0.05 level.

*3.3 Model Cross-validation*

The overall cross validation R2 for the entire study area and study period is 0.70 and the slope between predicted PM2.5 and the observed PM2.5 is 0.98, indicating a good agreement between CV estimates and observations in year 2011. The climate regional specific results of complete and spatial 10-fold cross validation including R2 and slope are listed in Table 4 and Table 5. Results show that the CV based performance of our model varies across different climate. For example, our model achieves the highest R2 under both cross-validation settings (complete R2=0.78, spatial R2=0.70) in the Northwest region, and in UpperMidwest and OhioValley regions the national downscaling model also have a strong prediction power in terms of complete R2 and spatial R2. On the other hand, the southwest region has the lowest R2 at 0.54 under complete 10-fold setting. Furthermore, under the spatial cross validation setting the model does not perform as good as the model under the complete cross validation setting where in the Northwest region the complete R2 is 0.60 and the spatial R2 is only 0.39. To be specific, Figure 3 and 4 show the scatterplot of CV estimates and the real PM2.5 concentration levels across nine climate regions. The CV based PM2.5 estimates have a good linear agreement with the real values in most climate regions including West, South, UpperMidwest, Southwest, Northwest and OhioValley. However, in Southwest, North and Northwest regions, the model tends to underestimate for higher PM2.5 concentrations.

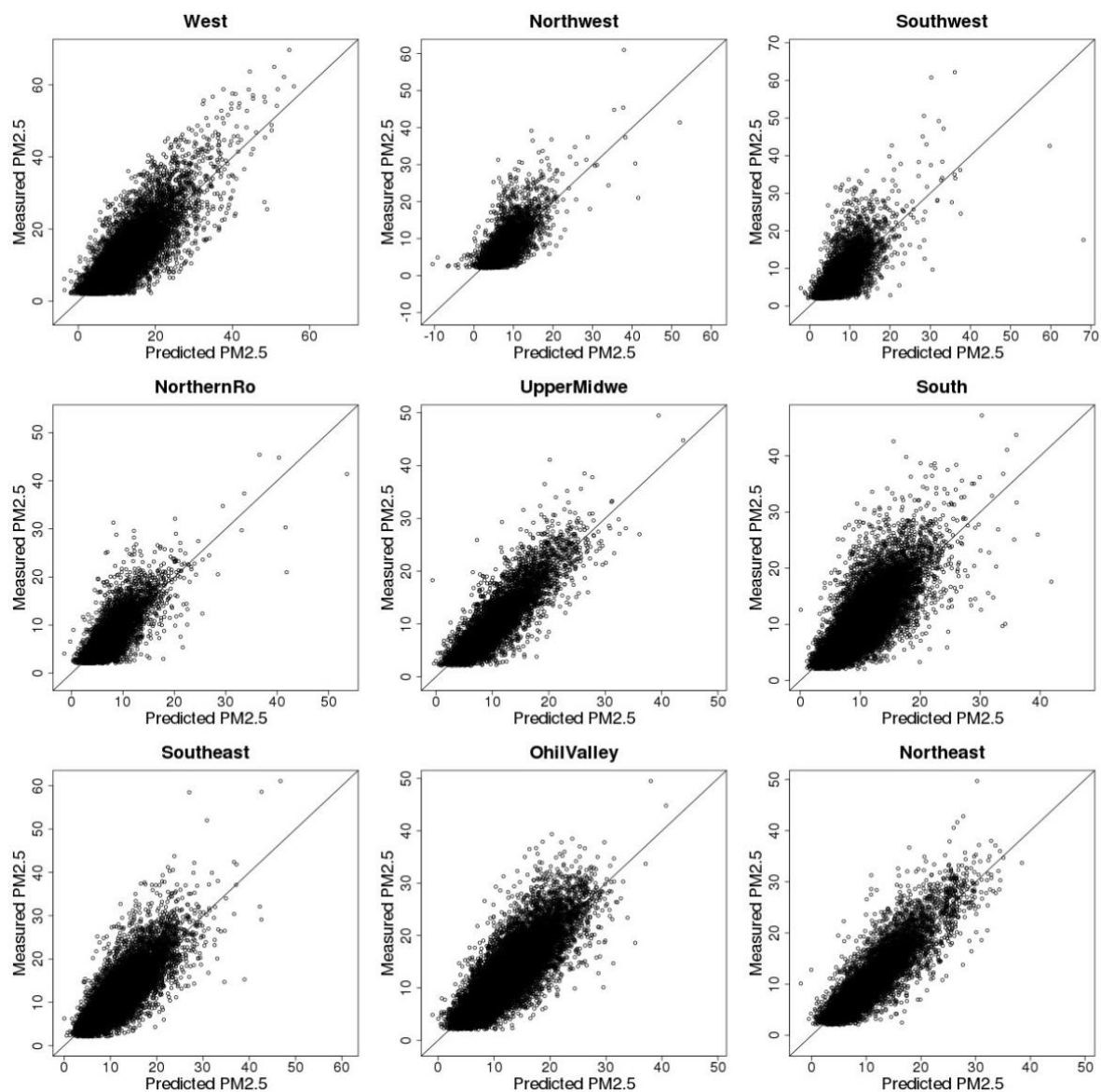

**Figure 3.** 10-fold cross validation results.

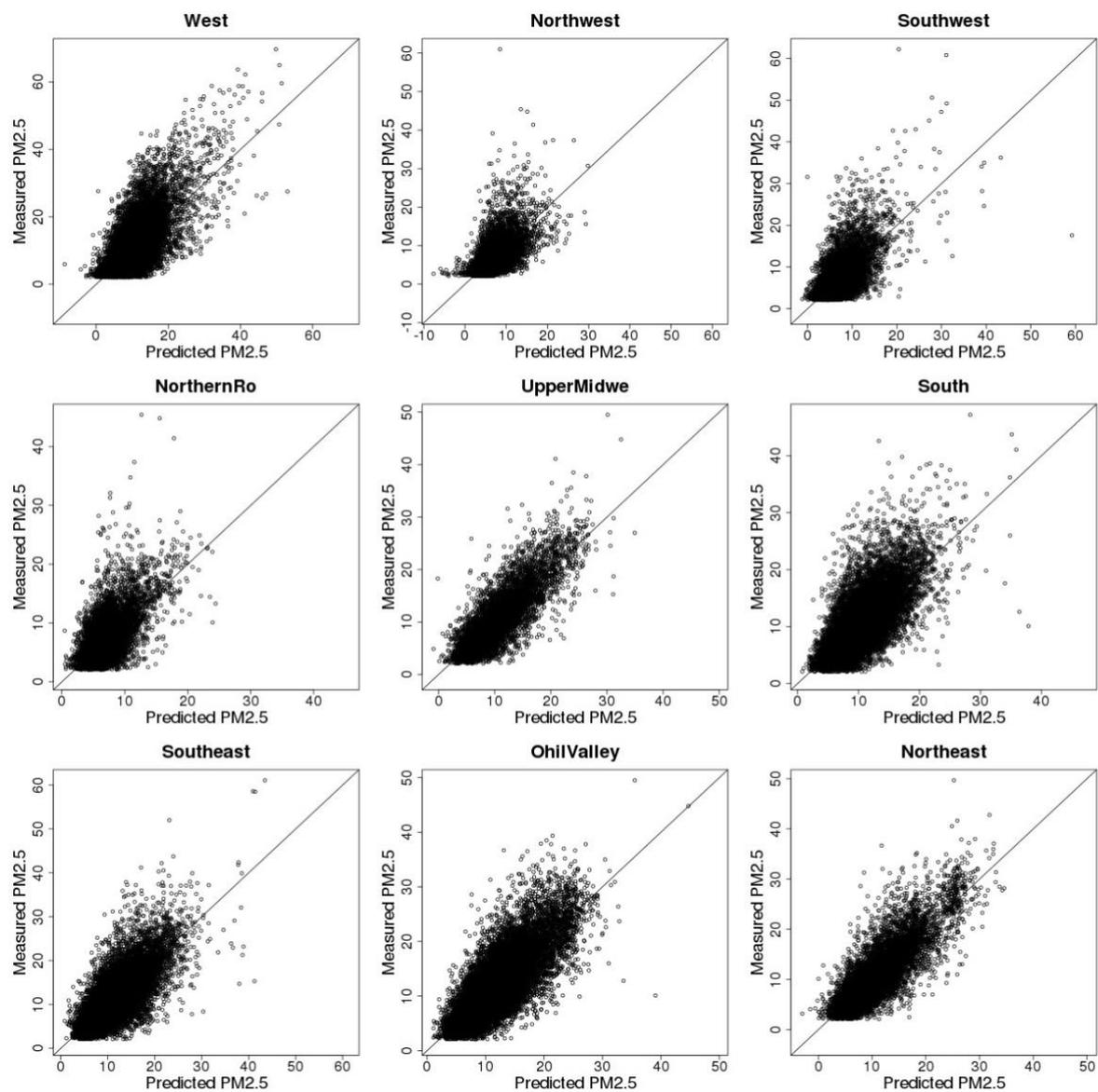

**Figure 4.** Spatial 10-fold cross validation results.

**Table 4.** 10-fold Cross Validation Results.

| Regions | R2 | Intercept | Slope |
|---|---|---|---|
| West | 0.69 | 0.04 | 0.99 |
| Northwest | 0.60 | 0.35 | 0.95 |
| Southwest | 0.54 | 0.40 | 0.94 |
| NorthernRo | 0.60 | 0.29 | 0.95 |
| UpperMidwe | 0.76 | −0.04 | 0.99 |
| South | 0.59 | 0.27 | 0.97 |
| Southeast | 0.69 | 0.19 | 0.98 |
| OhilValley | 0.71 | 0.07 | 0.99 |
| Northeast | 0.78 | 0.07 | 0.99 |

**Table 5.** Spatial 10-fold Cross Validation Results.

| Regions | R2 | Intercept | Slope |
|---|---|---|---|
| West | 0.46 | 0.36 | 1.02 |
| Northwest | 0.39 | 1.01 | 0.83 |
| Southwest | 0.40 | 0.96 | 0.87 |
| NorthernRo | 0.37 | 0.94 | 0.90 |
| UpperMidwe | 0.69 | -0.01 | 0.99 |
| South | 0.50 | 0.38 | 0.96 |
| Southeast | 0.58 | 0.77 | 0.92 |
| OhilValley | 0.65 | 0.18 | 0.97 |
| Northeast | 0.70 | 0.33 | 0.97 |

*3.4 Model Prediction*

Based on the simulated AOD from GEOS-Chem, we are able to prediction the PM2.5 concentration for the whole United States. The predicted annual average PM2.5 concentrations and its model based SD are visualized in Figure 5 and 6. As shown in Figure 5, the predicted annual mean of PM2.5 concentration is smoothed across all the spatial domains even among the climate buffer regions indicating that the national Bayesian model fits the data well. Furthermore, a strong spatial differential pattern exists in the annual PM2.5 spread where the PM2.5 concentration is higher in eastern regions than western regions. California, Great Lakes regions and the east coast regions including New York and Washington have an especially high annual average PM2.5 concentrations. On the other hand, the lowest annual PM2.5 concentration lies in the region of middle part of America including Utah, Colorado, Wyoming and Idaho which are the states having most forest coverage and least human behavior. These indicate that our model can capture the large scale spatial spread of PM2.5 very well. Model is also able to discover the small features of the predicted PM2.5 concentration surface, where we can observe high PM2.5 concentration levels in urban centers such as Atlanta, Dallas, Houston, Miami and Salt Lake City.

For the prediction uncertainty, Figure 6 shows the spatial spread of the standard deviation of the annual average PM2.5 concentrations. As shown in Figure 6, West region including California and Nevada has a higher SD compared with other regions. South and Southeast regions have lowest SD on average. More specifically, from the spatial distribution of SD, we also observe a higher peak at Miami, Houston and Dallas areas and Colorado state. Similarly, we also visualized the predicted seasonal average PM2.5 concentration and their SD and the results are in Supplemental Figure S1 and S2.

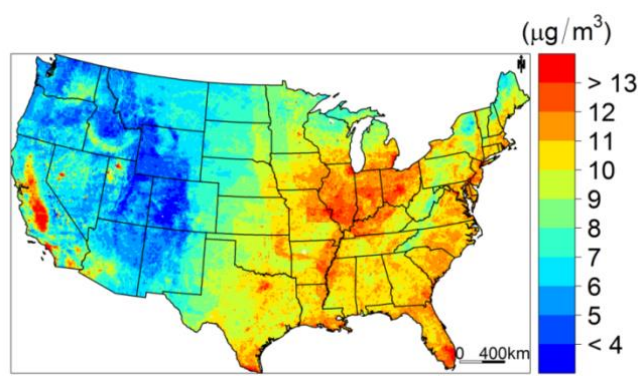

**Figure 5.** Predicted Annual PM2.5 Concentration across the Continental US.

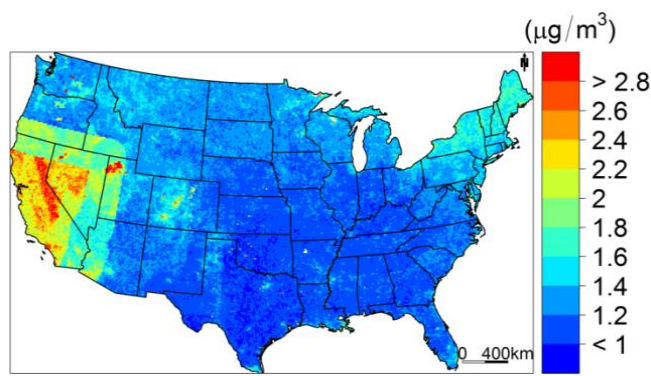

**Figure 6.** The Uncertainty (Standard Deviation) of the Predicted Annual PM2.5 Concentration across the Continental US.

## 4. Discussion and Conclusion

We presented a national Bayesian downscaler model to estimate daily PM2.5 concentrations in the Continental US using satellite aerosol remote sensing data, meteorological and land use parameters. Overall, our national Bayesian downscaling model performs well at the national scale. Compared with the regional models including regional hierarchical models, mixed effect model and regional Bayesian downscaling methods, our approach provides the nationally cohesive predictions and quantifies the model prediction errors very well. Compared with the machine learning models (e.g., neural networks and random forests), our approaches incorporate the core of the statistical approaches providing insights into the physical and geographical information of the problem and the model uncertainty provided by the Bayesian approach is much more informative than those generated from machine learning models. It has the advantage of explicitly displaying the important predictors of PM2.5 in different geographical regions, which allows model simplification and further improvements of model performance.

In our national Bayesian Downscaling approach, we first adopt the nine climate regions and three temporal regions to separate the data into sub-regions which provides more flexible model fitting and then we utilized the Bayesian downscaling approaches in each sub-blocks to quantify the geographical patterns and the association between AOD and PM2.5. Our approach has several strengths. First, our approach uses a latent spatial process to incorporate the spatial correlation which can borrow information across the neighborhoods and is able to make more reliable prediction compared with mixed effect models.

Second, based on the climate region and temporal separation, our model is much more flexible in the model fitting and therefore fits the data better than the traditional Bayesian models. As shown in the Table 1, there is a significant difference in the geographical patterns across regions and temporal domain and this difference revealed by the model is an important sign that in different climate region the association between AOD and PM2.5 is complex and regional-specific. This further confirms that using a single model for whole national domain is not realistic which cannot reveal the real physical mechanism researchers are interested at. Moreover, our proposed approach is able to parallel setting and is much faster than traditional approach.

Finally, compared with the machine learning approach for national calibration, although our approach has slightly weaker prediction ability, probably because of less predictors used in our models than in theirs. For example, Di et al. (2016) included more than 50 predictors in the neural network model, and Hu et al. (2017)'s random forest model contained ~40 predictors. In addition, both Di et al. (2016) and Hu et al. (2017) used convolutional layers for nearby PM2.5 measurements and land use terms in their models, and both studies point out that convolutional layers can help to improve prediction accuracy. Although we could include these predictors in our models, it will require additional computing resources and consume additional computing time. We will address this issue in future research. It should be noted that the goal of this approach is to study the geographical patterns across different regions and seasons and our approach successfully provides great insights into these and provides much more informative results than machine learning methods. As we mentioned, the limited prediction ability of our approach in some specific regions, i.e. South region, is one limitation. The reason for this is that even though we separate the national domain into sub-blocks, each region is still very large and is hard for a single model to fit across such large domain. Thus, one future direction of this approach is to provide more flexible approach.

**Author Contributions:** YL conceived this study. HC and LW contributed to study design. YW and XH performed model development and prepared the manuscript. JB processed model input data. All authors commented on the manuscript. Authorship must be limited to those who have contributed substantially to the work reported.

**Funding:** This work was partially supported by the NASA Applied Sciences Program (grant no. NNX16AQ28G; PI: Liu). This publication was developed under Assistance Agreement No. 83586901 awarded by the U.S. Environmental Protection Agency to Emory University (PI: Liu). It has not been formally reviewed by EPA. The views expressed in this document are solely those of [name of recipient or names of authors] and do not necessarily reflect those of the Agency. EPA does not endorse any products or commercial services mentioned in this publication.

**Data Availability Statement:** The datasets analyzed for this study can be found in the Google Drive [https://drive.google.com/drive/folders/1e4YQQM39_Qw6y6hwPZxjQ0uTTZB60DNU?usp=sharing].

**Conflicts of Interest:** The authors declare that the research was conducted in the absence of any commercial or financial relationships that could be construed as a potential conflict of interest.

**Appendix**

The appendix is an optional section that can contain details and data supplemental to the main text. For example, explanations of experimental details that would disrupt the flow of the main text, but nonetheless remain crucial to understanding and reproducing the research shown; figures of replicates for experiments of which representative data is shown in the main text can be added here if brief, or as Supplementary data. Mathematical proofs of results not central to the paper can be added as an appendix.